\documentclass[twocolumn,showpacs,preprintnumbers,amsmath,amssymb]{revtex4}

\usepackage{graphicx}
\usepackage{dcolumn}
\usepackage{bm}
\usepackage{epsfig}

\begin{document}


\title{Characterization of the non-classical nature of conditionally prepared single photons}

\author{Alfred B. U'Ren$^{1,2}$}
\author{Christine Silberhorn$^1$}
\author{Jonathan L. Ball$^1$}
\author{Konrad Banaszek$^1$}
\author{Ian A. Walmsley$^1$}
\affiliation{$^1$Clarendon Laboratory, Oxford University, Parks
Road,
Oxford, OX1 3PU, England \\
$^2$Centro de Investigaci\'{o}n Cient\'{i}fica y Educaci\'{o}n
Superior de Ensenada (CICESE), Baja California, 22860, Mexico}

\date{\today}
%
\newcommand{\epsfg}[2]{\centerline{\scalebox{#2}{\epsfbox{#1}}}}
\begin{abstract}
A reliable single photon source is a prerequisite for linear optical
quantum computation and for secure quantum key distribution. A
criterion yielding a conclusive test of the single photon character
of a given source, attainable with realistic detectors, is therefore
highly desirable. In the context of heralded single photon sources,
such a criterion should be sensitive to the effects of higher photon
number contributions, and to vacuum introduced through optical
losses, which tend to degrade source performance.  In this paper we
present, theoretically and experimentally, a criterion meeting the
above requirements.
\end{abstract}

\maketitle


High fidelity single photon sources are an essential ingredient for
quantum-enhanced technologies including linear optical quantum
computation (LOQC) and secure quantum key distribution.  Thus, the
endeavor to generate single photons in controlled, well-defined
spatio-temporal modes is an active area of research. Current single
photon source candidates can be classified into two categories:
deterministic sources producing single photons on demand at
predefined trigger times and heralded single photon sources relying
on the spontaneous emission of distinguishable photon pairs in
conjunction with conditional preparation. While the emission times
for conditional single photon sources cannot be controlled beyond
the restriction of emission time slots through a pulsed pump, it has
been shown that waveguided PDC can yield heralded single photons in
well defined modes together with high collection
efficiencies\cite{uren04}. Conditional state preparation has been
utilized in various physical systems including atomic cascades
\cite{grangier86}, ensembles of cold atoms \cite{chou04} and in
parametric downconversion (PDC).  In the case of PDC, conditional
preparation was first reported by Mandel \textit{et
al.}\cite{hong86} and since then has been optimized to generate
approximately true $n=1$ Fock states
\cite{uren04,rarity87,kwiat91,lvovsky01,alibart04,pittman04}. In
order to assess the performance of heralded single photon sources a
criterion that takes into account the detrimental contributions of
higher photon numbers and optical losses is needed. In addition,
such a criterion should take into full consideration limitations of
existing photodetectors such as the binary behavior of avalanche
photodiodes operated in the Geiger mode where a single click
signifies the detection of one or more photons.  In this paper we
derive such a criterion and show that our previously reported
waveguided PDC source\cite{uren04} represents a high fidelity source
of heralded single photons.

A standard approach used to determine whether a light source
exhibits classical or quantum photon statistics is the measurement
of a $g^{(2)}(\tau)$ second-order intensity autocorrelation function
in a Hanbury-Brown Twiss geometry. The semi-classical theory of
photodetection predicts, firstly, that $g^{(2)}(0) \geq
g^{(2)}(\tau)$ for all time delays $\tau$, and, secondly, that
$g^{(2)}(0) \geq 1$. The observation of photon anti-bunching,
\textsl{i.e.} $g^{(2)}(0) \leq g^{(2)}(\tau)$, has been utilized,
for example, to verify the non-classical character of deterministic
single photon sources implemented by strongly coupled atom cavity
systems \cite{mcKeever04}. For PDC sources, the probability of
generating simultaneously two photon pairs at a given instant of
time is of the same order as the probability of generating two
independent pairs separated by the interval $\tau$. This obliterates
the effect of antibunching, unless we employ selective heralding
that identifies specifically a single-pair component. For PDC
sources the non-classical character of the generated radiation is
usually tested by violating the lower bound on the second-order
intensity autocorrelation function $g^{(2)}(0) \geq 1$. The value of
$g^{(2)}(0)$ constitutes a figure of merit which determines the
degree to which higher photon number contributions degrade the
single photon character \cite{alibart04}.

\begin{figure}[ht]
\epsfg{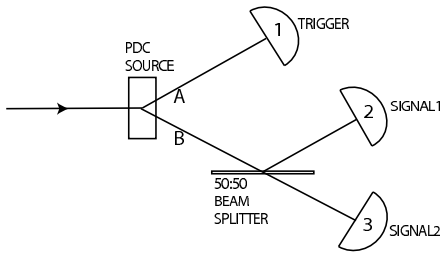}{.6} \caption{Schematic of experimental
setup designed to test the single photon character of a
conditionally prepared state} \label{Fi:BBineqSchematic}
\end{figure}

Based on a classical wave description and intensity measurements,
Grangier \textsl{et al.} derived from the Cauchy Schwarz inequality
a similar ``anti-correlation'' criterion for characterizing
conditionally prepared single photons by coincidence detection rates
\cite{grangier86}. For the experimental configuration shown in
Fig.~\ref{Fi:BBineqSchematic} an anti-correlation parameter:
\begin{equation}
\alpha=\frac{R_1 R_{123} }{R_{12} R_{13}}
\end{equation}
can be defined, which indicates non-classical photon statistics for
$\alpha <1$, where $R_i$ represents the singles count rates at
detector $i$, and $R_{ij}$, $R_{ijk}$ the double and triple
coincidences for the respective detectors $i,j,k$. A variant of the
$g^{(2)}(0)$ measurement specifically designed to study conditional
single photon sources independently from losses, which has been
pioneered by Clauser \cite{Clauser74}, has recently been implemented
for single photons generated from an ensemble of cold atoms
\cite{chou04}.

In the above works, the theoretical modeling of experimental data
was carried out in terms of intensity correlation functions. In a
typical experiment, however, the count rates are directly related to
light intensities only under certain auxiliary assumptions. The
reason for this is that standard photodetectors sensitive to single
photons, such as avalanche photodiodes operated in the Geiger
regime, do not resolve multiphoton absorption events and yield only
a binary response telling us whether at least one photon was present
in the detected mode or none at all. With such detectors, the light
intensity can be read out from the count rates only in the limit of
weak fields, where the probability of detecting a single photon is
proportional to the intensity. In a general case, the probability of
obtaining a click is a nonlinear function of the incident intensity.
This aspect is particularly important in schemes utilizing
ultrashort pulses, where the incoming light energy is concentrated
in sub-picosecond time intervals that cannot be resolved even by the
fastest photodetectors. It is therefore interesting to go beyond the
basic intensity correlation theory and examine whether count
statistics collected with binary non-photon-resolving detectors can
serve as a test of source non-classicality. We will demonstrate in
the following that this is indeed the case. Furthermore, the
non-classicality criterion based on measuring $g^{(2)}(0)$ relies on
a coincidence basis measurement so that losses can be neglected.
However, for applications such as cascaded logic gates in
LOQC\cite{knill01} and loophole free tests of Bell
inequalities\cite{kwiat95} post-selection is not desirable, as it
leads to vacuum contamination. The latter diminishes the usability
of the single photon states: heralding no longer necessarily
corresponds to the successful generation of a single-photon or LOQC
gate operation. In this paper we derive a criterion designed to test
the non-classical nature of conditionally prepared single photon
states. Our criterion takes into account both, the non-linearity of
the detectors and the fidelity of the generated single-photon state,
which measures the probability that a single-photon is actually
present when it is heralded. The criterion can be tested in a
standard setup in which the signal field is subdivided into two
submodes, each monitored by a non-photon number resolving detector.

Consider a source emitting two light beams whose intensities,
integrated over the detector active area, are $W_A$ and $W_B$. In
the semiclassical theory of photodetection we will treat these
intensities as positive-definite stochastic variables described by a
joint probability distribution ${\cal P}(W_A;W_B)$. Beam $B$ is
divided by a beam splitter with power reflection and transmission
coefficients $r$ and $t$. Finally, the resulting beams are detected
by three photodetectors. We will assume that the probability of
obtaining a click on the $i$th detector illuminated by intensity $W$
is given by $p_i(W)$, bounded between $0$ and $1$. We furthermore
assume $p_i(W)$ to be a monotonic increasing function of its
argument $W$. Under these assumptions it is easy to show that the
following inequality is satisfied for an arbitrary pair of arguments
$W_B$ and $W'_B$:
\begin{equation}
[p_2(rW_B) - p_2(rW'_B)][p_3(tW_B) - p_3(tW_B')] \ge 0
\end{equation}
Indeed, the sign of both the factors in square brackets is always
the same, depending on the sign of the difference $W_B-W'_B$; their
product is therefore never negative. Let us now multiply both sides
of the above inequality by the factor ${\cal P}(W_A;W_B){\cal
P}(W_A';W_B')p_1(W_A)p_1(W'_A)$ which is likewise nonegative, and
perform a double integral $\int_0^\infty dW_A dW_B \int_0^\infty
dW_A' dW_B'$. This yields the inequality:
\begin{equation}
\label{R1R123-R12E13}B= R_1 R_{123} - R_{12} R_{13} \ge 0
\end{equation}
where the single, double, and triple count rates are given by
averages $\langle \ldots \rangle = \int_0^\infty dW_A dW_B {\cal
P}(W_A;W_B) \ldots$ defined with respect to the probability
distribution ${\cal P}(W_A;W_B)$:
\begin{eqnarray*}
R_1     & = & \langle p_1(W_A) \rangle \\
R_{12}  & = & \langle p_1(W_A) p_2(rW_B) \rangle \\
R_{13}  & = & \langle p_1(W_A) p_3(tW_B) \rangle \\
R_{123} & = & \langle p_1(W_A) p_2(rW_B) p_3(tW_B) \rangle
\end{eqnarray*}
It is seen that the inequality derived in Eq.~(\ref{R1R123-R12E13})
which can be transformed into:
\begin{equation}
\frac{R_1 R_{123}}{R_{12} R_{13}} \ge 1
\end{equation}
has formally the same structure as the condition derived by Grangier
{\em et al.}\cite{grangier86}. However, the meaning of the count
rates is different, as we have incorporated the binary response of
realistic detectors. It is noteworthy that this inequality has been
derived with a very general model of a detector, assuming
essentially only a monotonic response with increasing light
intensity.

Our experimental apparatus is similar to that reported in
Ref.~\cite{uren04}. PDC is generated by a KTP nonlinear waveguide
pumped by femtosecond pulses from a modelocked, frequency doubled
87MHz repetition rate Ti:sapphire laser. In contrast to that
reported in Ref.~\cite{uren04}, the approach here is to record
time-resolved detection information for the three spatial modes
involved with respect to the Ti:sapphire pulse train as detected by
a fast photodiode. We thus obtain a reference clock signal with
respect to which post-detection event selection can be performed in
order to implement temporal gating.  The latter is important for the
suppression of uncorrelated background photons, the presence of
which can lead to heralded vacuum (rather than a true single
photon). Through this approach, we are able to freely specify the
time-gating characteristics; arbitrarily complicated logic can be
performed without added experimental hardware. Drawbacks include the
lack of real-time data processing as well as the deadtime in the
region of $\mu$s between subsequent triggers exhibited by the
digital oscilloscope (LeCroy WavePro 7100) used for data
acquisition. In our setup, source brightness information is obtained
via a separate NIM electronics-based measurement.

\begin{figure}[ht]
\epsfg{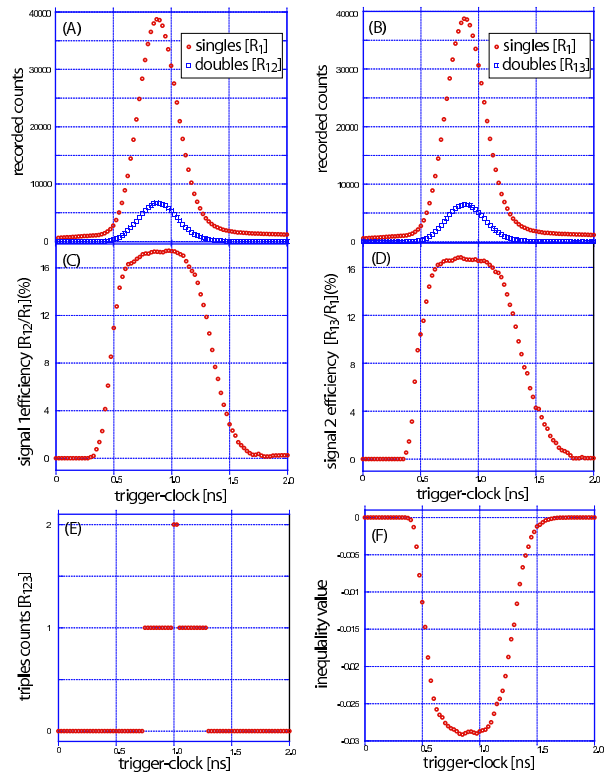}{.5} \caption{(color online) (a)
Time-resolved doubles and singles count rates for signal$_1$.  (b)
Time-resolved doubles and singles count rates for signal$_2$.  (c)
Conditional detection efficiency for signal$_1$. (d) Conditional
detection efficiency for signal$_2$. (e) Time-resolved triple count
rate (f) Time-resolved inequality parameter exhibiting violation.}
\label{Fig:GrangierData}
\end{figure}

For a given trigger event, three numbers are recorded: the time
difference between the electronic pulse positive edge corresponding
to the trigger and to the two signal modes $t_{S1}$, $t_{S2}$, as
well as the trigger-clock reference time difference $t_{CLK}$.
Time-gating involves discarding trigger events outside a certain
range of $t_{CLK}$ values, while coincidence events with $t_{S1}$
and $t_{S2}$ outside a $1.1$ns wide coincidence window are regarded
as accidental and ignored. We collected $75000$ trigger events and
measured pre-time gating detection efficiencies (defined as the rate
of coincidences normalized by singles) for each of the two signal
channels of $14.4\%$ and $13.7\%$. Fig.~\ref{Fig:GrangierData} shows
the post-processed data using a scanned temporal band-pass filter
with $300$ps width (selected to approximately match the measured APD
jitter). Fig.~\ref{Fig:GrangierData}(A)[(B)] shows the time-resolved
signal$_1$-trigger [signal$_2$-trigger] coincidence count rate,
compared to the time-resolved trigger singles count rate.
Fig.~\ref{Fig:GrangierData}(C)[(D)] shows the resulting time-gated
detection efficiency for the signal$_1$ [signal$_2$] channel,
showing maximum values of $\sim17.4\%$ [$\sim17.0\%$].
Fig.~\ref{Fig:GrangierData}(E) shows time-resolved triple
coincidences, for identical coincidence windows as used in computing
double coincidences. Thus, our time-gating procedure filters the PDC
flux so that for the pump-power used the generated light is
described essentially by a superposition of vacuum with single
photon pairs, showing nearly vanishing multiple pair generation.

Fig.~\ref{Fig:GrangierData}(F) shows the time-resolved inequality
parameter [see Eq.~\ref{R1R123-R12E13}] resulting from the count
rates presented above. As a numerical example, at the peak of the
triples counts, we obtain the following time-gated counting rates
for $75000$ trigger events: $R_{123}=2$, $R_{12}=5329$,
$R_{13}=5067$ and $R_1=30629$, yielding an inequality parameter
value of $B=-0.029 \pm .001$. For comparison, our results correspond
to a value of the anti-correlation parameter of $\alpha= (2.3 \pm
1.6)\times 10^{-3}$. Fig.~\ref{Fig:GrangierData} indicates an
overall signal transmission [defined as the sum of the two
individual efficiencies $(R_{12}+R_{13})/R_1$] of $\sim$34.5$\%$.
The main contribution to losses is the non-unit quantum efficiency
of the single photon detectors.  The overall detection efficiency is
also degraded due to imperfect optical transmission and remaining
unsuppressed uncorrelated photons.   From the above count rates, we
can also calculate $g^{(2)}(0)=2 p_{(2)}/p_{(1)}^2$ in terms of the
probability of observing a single photon in the signal arm
$p_{(1)}=(R_{12}+R_{13})/R_1$ and the probability of observing two
photons in the signal arm $p_{(2)}=R_{123}/R_1$. We thus obtain
$g^{(2)}(0)=(1.1\pm 0.8)\times 10^{-3}$, amongst the lowest reported
for a single photon source.

\begin{figure}[ht]
\epsfg{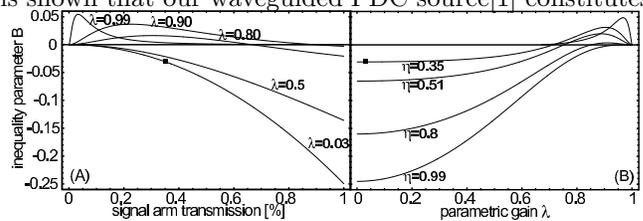}{.55} \caption{(A)Inequality parameter $B$ as a
function of the overall transmission in the signal arm (for fixed
$\lambda$ values as shown). (B) Inequality parameter $B$ as a
function of the parametric gain (for fixed $\eta$ values as shown).
Black squares represent the experimentally obtained value for the
inequality parameter assuming $\lambda=0.03$; the uncertainty is
smaller than the square dimensions.  A trigger detection of
efficiency $\eta_{T}=0.02$ was assumed.} \label{Fig:BBtheo}
\end{figure}

Ignoring the spectral and transverse momentum degrees of freedom,
the signal and idler photon-number distribution in a realistic PDC
source is expressed as:
\begin{equation}
|\Psi\rangle=\sqrt{1-|\lambda|^2}\sum\limits_{n=0}^\infty
\lambda^n|n\rangle_s|n\rangle_i
\end{equation}
where $n$ represents the photon number describing each of the signal
and idler modes and $\lambda$ represents the parametric gain. PDC
experiments often operate in a regime where $\lambda$ is small
enough that the probability of multiple pair generation becomes
negligible. For larger values of $\lambda$ (accessed for example by
a higher pump power or higher non-linearity), however, the higher
order terms (e.g. $|2\rangle_s|2\rangle_i$,
$|3\rangle_s|3\rangle_i$...) become important. While these higher
photon number terms are desirable for conditional preparation via
photon number resolving detection, in the context of the present
work, where the detectors used \textit{are not} photon-number
resolving and where the emphasis is on high-fidelity preparation of
\textit{single} photons, multiple pair generation must be avoided.
As discussed earlier,in order to characterize a source of
conditionally prepared single photons based on PDC, besides the
parametric gain $\lambda$, optical losses must be taken into
account.  Losses in the signal arm imply that a trigger detection
event can incorrectly indicate the existence of a signal photon,
while in reality vacuum is present. Fig.~\ref{Fig:BBtheo} shows the
expected inequality behavior based on a quantum mechanical
calculation in which it is assumed that the detection probability is
given by the expectation value of the operator $1-\exp(-\eta
\hat{W})$ (where $\hat{W}$ is the time-integrated incoming intensity
operator and $\eta$ is the corresponding overall transmission
including all optical and detection losses).
Fig.~\ref{Fig:BBtheo}(A) shows the calculated inequality parameter
$B$ for PDC light as a function of the overall signal optical
transmission $\eta_s=(R_{12}+R_{13})/R_1$ for a fixed value of the
parametric gain $\lambda$. Fig.~\ref{Fig:BBtheo}(B) shows the
inequality coefficient as a function of the parametric gain
$\lambda$ for different levels of optical loss.  Note that a strong
violation of the inequality is only observed in the low parametric
gain limit coupled with low losses.  Note further that the minimum
value of $B$, corresponding to the strongest violation and which is
only reached in the ideal lossless case, is $-0.25$. In an
experimental realization, while accessing very low values of
$\lambda$ is straightforward \textit{e.g.} by using a low pump
power, attaining a sufficiently low level of loss to yield a nearly
ideal violation is challenging. An analysis of expected detection
rates, under the assumption that all uncorrelated photons in the
trigger arm are suppressed, yields the parametric gain $\lambda$ in
terms of experimentally measurable quantities:
\begin{equation}
\lambda^2=\frac{R_2+R_3}{\eta_s R_{rep} (1+f)}
\end{equation}
where $R_{rep}$ is the pump repetition rate and $f$ is the
uncorrelated photon intensity normalized by that of PDC. We estimate
that in our experiment $f$ is constrained by: $0<f\lesssim 2$. Our
experimental values of $R_2+R_3\approx 70000 s^{-1}$,
$R_{rep}=87\times10^6 s^{-1}$ and $\eta_s=0.345$ thus yield:
$0.016<\lambda<0.047$. The experimentally observed violation [see
Fig.~\ref{Fig:GrangierData}(F)] is in good agreement with the theory
curves in  Fig.~\ref{Fig:BBtheo}. The black squares in
Fig.~\ref{Fig:BBtheo}(A) and (B) depict the observed violation as
compared with the theoretical curves, where the uncertainty is
smaller than the square dimensions. The plot in
Fig.~\ref{Fig:BBtheo}(A) assumes a fixed value of the parametric
gain $\lambda$ (with different curves shown for a choice of
$\lambda$ values). The signal arm transmission is obtained as the
sum of the two individual signal detection efficiencies [see
Fig.~\ref{Fig:GrangierData}(A) and (B)].

In summary, we have derived a criterion which allows a conclusive
test of the single photon character of conditionally prepared single
photon states.  We have shown that the inequality in
Eq.~\ref{R1R123-R12E13} is fulfilled by all classical light sources,
as well as by states generated by PDC exhibiting higher photon
numbers through a large parametric gain. On the contrary, a strong
violation of the inequality is observed only for states that
constitute a good approximation to a conditionally prepared single
photon.  Our criterion is realistic enough to include binary
non-photon number resolving photon counting detectors while it is
sensitive to the degradation observed in the prepared state caused
by a vacuum component due to losses, crucial for assessing heralded
single photon source performance. Through the application of our
criterion it is shown that our waveguided PDC source\cite{uren04}
constitutes a high-fidelity conditional single photon source. Our
derived inequality yields a new figure of merit quantifying the
overall performance of conditional single photon sources taking into
full consideration experimental imperfections.


\end{document}